\documentclass[11pt,a4paper]{article}
\usepackage{amsfonts}  

\newcommand{\1}{\mathbb{I}}
\newcommand{\C}{\mathbb{C}}

\newcommand{\Z}{\mathbb{Z}}
\newcommand{\G}{\Gamma_n}
\newcommand{\cG}{-\!\!\!\Gamma_n}
\renewcommand{\L}{\mbox{L}^2}
\newcommand{\ran}{\mbox{ran}}
\newcommand{\mod}{\mbox{ mod }}
\newcommand{\nulty}{\mbox{null}}

\title{Two particles on a star graph II}
\author{Mark Harmer\\
email: harmer@ujf.cas.cz\\
Czech Academy of Sciences\\
\v{R}e\v{z}\\
Czech Republic}

\begin{document}

\maketitle 

\begin{abstract}
We consider a two particle system on a star graph with $\delta$-function interaction. A complete description of the eigensolutions with real momenta is given; specifically it is shown that all eigensolutions can be written as integrals in the momentum plane of sums of products of appropriate one particle solutions.
\end{abstract}

\section{Introduction and Summary}
This paper is a continuation of \cite{Har9} in which was considered the two particle problem with $\delta$-function interaction on a star graph with $n$ edges, $\G$. We briefly recall the problem: the configuration space is $\G^2 = \G \times \G$ with the local coordinate neighbourhoods consisting of $n^2$ quadrants $\left\{ Q_{ij} = \{ (x_i, y_j) \} \right\}^{n}_{i,j=1}$. The hamiltonian is defined as
\begin{equation}\label{Hc}
\left. H_c \right|_{Q_{ij}} = \left. \Delta\right|_{Q_{ij}} + c\, \delta_{ij}\, \delta(x_i - y_i)
\end{equation}
with $c$ real, $\left. \Delta\right|_{Q_{ij}} = \mbox{} - \frac{\partial^2}{\partial^2 x_i} - \frac{\partial^2}{\partial^2 y_j}$, and Kirchhoff boundary conditions at the boundaries of the quadrants
\begin{eqnarray}
\left. \psi \right|_{Q_{ij}, x_{i}=0} = \left. \psi \right|_{Q_{kj}, x_{k}=0} \; ; \quad \sum^n_{l=1} \left. \frac{\partial \psi}{\partial x_l} \right|_{Q_{lj}, x_{l}=0} = 0 \, , \; \forall i, j, k \label{bbc1} \\
\left. \psi \right|_{Q_{ij}, y_{j}=0} = \left. \psi \right|_{Q_{ik}, y_{k}=0} \; ; \quad \sum^n_{l=1} \left. \frac{\partial \psi}{\partial y_l} \right|_{Q_{il}, y_{l}=0} = 0 \, , \; \forall i, j, k \, . \label{bbc2}
\end{eqnarray}
The $\delta$ interaction on the diagonal $D=\left\{ x_i=y_i \, ; \, x_i,y_i\in Q_{ii}, i\in\{1,\ldots ,n \} \right\}$ means that eigensolutions satisfy the boundary conditions \cite{Alb:Kur}
\begin{eqnarray}
& \left. \psi \right|_{Q_{ii}, x_i = y^+_i} = \left. \psi \right|_{Q_{ii}, x_i = y^-_i} & \label{dbc1} \\
& \left. \frac{1}{2} \left( \frac{\partial \psi}{\partial x_i} - \frac{\partial \psi}{\partial y_i} \right) \right|_{Q_{ii}, x_i = y^+_i} - \left. \frac{1}{2} \left( \frac{\partial \psi}{\partial x_i} - \frac{\partial \psi}{\partial y_i} \right) \right|_{Q_{ii}, x_i = y^-_i} = c\cdot \left. \psi \right|_{Q_{ii}, x_i = y_i} \, . & \label{dbc2}
\end{eqnarray}
This problem (\ref{Hc}) is interesting because the equivalent problem on the line ($n=2$) is `completely solvable', not just for two particles but for an arbitrary number of particles \cite{Yan}. In this paper we use the results from \cite{Har9} and describe {\em all} solutions to the eigenvalue problem for $H_c$ with real momenta. To be precise we describe all eigensolutions of $H_c$ of energy $\lambda>0$ which can be written in the form
\begin{eqnarray}
\psi (x,y) & = & \int_{\sqrt{\lambda}I} \left( \mbox{} - e^{ikx + i\sqrt{\lambda-k^2}y}\, \psi^{++} + e^{-ikx + i\sqrt{\lambda-k^2}y}\, \psi^{-+} \right. \nonumber \\
& & \left. \mbox{} - e^{-ikx - i\sqrt{\lambda-k^2}y}\, \psi^{--} + e^{ikx - i\sqrt{\lambda-k^2}y}\, \psi^{+-} \right) dk \label{bsoln}
\end{eqnarray}
on the off diagonal quadrants, or on one of the sectors $\{x_i>y_i\}$, $\{x_i<y_i\}$ of the diagonal quadrants. Here $I=[0,1]$ and since
$$
\lim_{R\to\infty} \frac{1}{R} \int^R_0 \int^R_0 | \psi |^2 \, dx \, dy = 2 \pi \sum_{\sigma,\tau} \int_{\sqrt{\lambda}I} | \psi^{\sigma \tau} |^2 dk
$$
we assume that the transforms $\psi^{\sigma \tau}(k)$ are square integrable on $\sqrt{\lambda}I$. We use the notation $\psi^{\sigma \tau}_{ij}$ to denote the transform on the quadrant $Q_{ij}$. \\
In this paper we show that these (real momentum) eigensolutions of $H_c$ call all be written as integrals in the momentum plane of the $2n^2-2n$ eigensolutions of $H_c$ described in \S4 of \cite{Har9}. These solutions were characterised by the fact that they are finite sums of products of one particle solutions on the configuration space cut along the diagonal, $\cG^2$. Let us denote here this $2n^2-2n$-dimensional vector space of solutions by $W$ and the individual solutions by 
$$
G^i \left( k_1, k_2 \right) \, .
$$
In order to prove this we first consider solutions (\ref{bsoln}) satisfying conditions (\ref{bbc1}, \ref{bbc2}) at the boundaries of the quadrants but which are not smooth on the diagonal and which did not, necessarily, satisfy (\ref{dbc1}, \ref{dbc2}). For want of a better term we refer to such solutions as `basic solutions'. It is shown in \S\ref{defel} that (\ref{bbc1}, \ref{bbc2}) imposes linear conditions on appropriate vectors of transforms $\psi^{\sigma \tau}$. Using these linear conditions we see that all `basic solutions' can be written as integrals (in the momentum plane) of the $n^2+n+1$ solutions described in \S3 of \cite{Har9}. In order to be precise and introduce some notation let us denote the solutions described in \S3 of \cite{Har9} by
$$
F^i \left( k_1, k_2 \right) 
$$
and the $n^2+n+1$-dimensional vector space they span by $V$. Then it is shown in \S\ref{defel} that the basic solutions correspond to $\L \left( \sqrt{\lambda}I \right)\otimes V$ under the map
\begin{equation}\label{ontodef}
\sum_i \int_{\sqrt{\lambda}I} f_i (k) \, F^i \left( k , \sqrt{\lambda-k^2} \right) dk 
\end{equation}
with $(f_i)\in\L \left( \sqrt{\lambda} I \right)\otimes V$. \\
In \S\ref{EHc} we consider the imposition of the boundary conditions (\ref{dbc1}, \ref{dbc2}) on the basic solutions. These impose {\em linear} conditions on the transforms, but {\em only} if we consider the permutation of the one particle momenta, ie. we need to consider transforms at momenta $k$ and $\sqrt{1-k^2}$, see (\ref{trnfd}); or equivalently $F^i \left( k , \sqrt{\lambda-k^2} \right)$ and $F^i \left( \sqrt{\lambda-k^2}, k \right)$; or equivalently $F^i \left( k_1 , k_2 \right)$ and $F^i \left( k_2 , k_1 \right)$. On the other hand, in \S4 of \cite{Har9} all eigensolutions of $H_c$ (constructed as products of one particle solutions) are found by enumerating those {\em linear} combinations of $F^i \left( k_1 , k_2 \right)$ and $F^i \left( k_2 , k_1 \right)$ which satisfy the boundary conditions (\ref{dbc1}, \ref{dbc2}). But this is exactly the problem described above from \S\ref{EHc}. In this way we see that the set of all solutions, at given energy $\lambda>0$ and with real momenta, to the eigenvalue problem for $H_c$ is simply the tensor product $\L \left( \sqrt{\frac{\lambda}{2}} I \right)\otimes W$. The map from the tensor product to the eigensolutions of $H_c$ is 
\begin{equation}\label{ontosoln}
\sum_i \int_{\sqrt{\frac{\lambda}{2}} I} g_i (k) \, G^i \left( k , \sqrt{\lambda-k^2} \right) dk 
\end{equation}
with $(g_i)\in\L \left( \sqrt{\frac{\lambda}{2}} I \right)\otimes W$. \\
In the last part of \S\ref{EHc} we recall the form of the $2n^2 - 2n$ eigensolutions $G^i$. We present them here in a slightly more geometric `nicer' form than in \cite{Har9}. 
\section{Description of the Basic Solutions}\label{defel}
Since the boundary conditions (\ref{bbc1}, \ref{bbc2}) are scale invariant we may rescale $x_i$ and $y_i$ on $\G^2$ thereby normalising $\lambda\equiv1$---in the following sections we put $\lambda\equiv1$ in (\ref{bsoln}, \ref{ontosoln}, \ref{ontodef}). From (\ref{bsoln}) 
\begin{equation}
\psi(x,0) = \int_{I} \left( e^{ikx} \left[ \mbox{} - \psi_{++} + \psi_{+-} \right] + e^{-ikx} \left[ \psi_{-+} - \psi_{--} \right] \right) dk
\end{equation}
and 
\begin{equation}
\frac{\partial \psi(x,0)}{\partial y} = \int_{I} i\sqrt{1-k^2} \left( e^{ikx} \left[ \mbox{} - \psi_{++} - \psi_{+-} \right] + e^{-ikx} \left[ \psi_{-+} + \psi_{--} \right] \right) dk 
\end{equation}
so that the Kirchhhoff boundary conditions (\ref{bbc2}) are written
\begin{eqnarray*}
& \sum_{l} \left( \psi^{++}_{il} + \psi^{+-}_{il} \right) = 0 = \sum_{l} \left( \psi^{--}_{il} + \psi^{-+}_{il} \right) & \\
&\psi^{++}_{ij} - \psi^{+-}_{ij} = \psi^{++}_{ik} - \psi^{+-}_{ik} \;\; , \;\; \psi^{--}_{ij} - \psi^{-+}_{ij} = \psi^{--}_{ik} - \psi^{-+}_{ik} &
\end{eqnarray*}
for all $i,j,k$. Similarly, the Kirchhhoff boundary conditions (\ref{bbc1}) take the form
\begin{eqnarray*}
& \sum_{l} \left( \psi^{++}_{lj} + \psi^{-+}_{lj} \right) = 0 = \sum_{l} \left( \psi^{--}_{lj} + \psi^{+-}_{lj} \right) & \\
&\psi^{++}_{ij} - \psi^{-+}_{ij} = \psi^{++}_{kj} - \psi^{-+}_{kj} \;\; , \;\; \psi^{--}_{ij} - \psi^{+-}_{ij} = \psi^{--}_{kj} - \psi^{+-}_{kj} &
\end{eqnarray*}
for all $i,j,k$. These can be written in compact form by introducing $\xi,\chi\in M_n(\C^2)$ defined as
\begin{equation}\label{trnf}
\xi_{ij} = \left( \begin{array}{c}
\psi^{++}_{ij} \\ \psi^{--}_{ij} \end{array} \right)  \; , \; 
\chi_{ij} = \left( \begin{array}{c}
\psi^{+-}_{ij} \\ \psi^{-+}_{ij} \end{array} \right) \, .
\end{equation}
Then (\ref{bbc2}, \ref{bbc1}) can be written as
\begin{equation}\label{bbc3}
\xi_{ij} = \mbox{} - \sum_{k} \chi_{ik} S^k_j \; , \; \xi_{ij} = \mbox{} - \sum_{k} S^k_i \tau \chi_{kj} \, .
\end{equation}
where $S^l_{j}$ is the unitary matrix $2P-\1$ with $P$ the projection onto $(1,1, \ldots ,1)^t$ and $\tau$ is the transposition on $\C^2$. \\
We denote by $\hat{\psi}^{\sigma \tau}_{ij}$ the transforms which, on the diagonal quadrants, give the eigenfunction in the sector $x_i>y_i$. Similarly we denote by $\check{\psi}^{\sigma \tau}_{ij}$ the transforms which, on the diagonal quadrants, give the eigenfunction in the sector $x_i<y_i$; in particular $\hat{\psi}^{\sigma \tau}_{ij}=\check{\psi}^{\sigma \tau}_{ij}$ for $i\ne j$. We define $\hat{\xi}$, $\hat{\chi}$ and $\check{\xi}$, $\check{\chi}$ similarly. Then using (\ref{bbc3}) we see that all solutions to the two particle problem satisfying (\ref{bbc1}, \ref{bbc2}) (but not necessarily (\ref{dbc1}, \ref{dbc2})) can be described by the system
\begin{equation}\label{defs}
\pi^{\perp}\left( \hat{\chi} S - S \tau \check{\chi} \right) = 0 \; , \; \pi^{\perp}\left( \hat{\chi} - \check{\chi} \right) = 0
\end{equation}
where $\pi$ is the projection onto the diagonal in $M_n$. This equation is finite dimensional, and linear, implying that all basic solutions can be described by a finite dimensional vector space (tensored by a space of functions). \\
It is shown in the appendix that the solutions of (\ref{defs}) can be split into four subspaces of dimension $(n-1)^2+1$, $2(n-1)$, $n-1$ and 2 respectively. Likewise the vector space $V$ of solutions, from \S3 of \cite{Har9}, splits into four: the $(n-1)^2+1$-dimensional smooth, symmetric subbasis; the $2(n-1)$-dimensional smooth, antisymmetric subbasis; the $n-1$-dimensional non smooth, symmetric subbasis; and the 2-dimensional non smooth, antisymmetric subbasis. It is a straightforward, but long, calculation to show that under (\ref{ontodef}) these four subbases of $V$ are mapped onto the solution subspaces of (\ref{defs}) of the same dimension. Consequently, the space of basic solutions is spanned by solutions of the form (\ref{ontodef}). 
\section{Eigensolutions of $H_c$}\label{EHc}
The eigensolutions of $H_c$ may be defined as the subspace of basic solutions which satisfy the boundary conditions (\ref{dbc1}, \ref{dbc2}) on $D$. In \S\ref{EHc1} we continue the above analysis describing the eigensolutions in terms of the transforms $\hat{\psi}^{\sigma\tau}$, $\check{\psi}^{\sigma\tau}$ and showing that they are given by (\ref{ontosoln}). In \S\ref{EHc2} we rewrite the eigensolutions described in \cite{Har9} (the $G^i$) presenting them in a slightly more geometric way then in \cite{Har9}.
\subsection{Boundary conditions on transforms on the diagonal}\label{EHc1}
The main point of this subsection is that, in analogy with (\ref{defs}), the boundary conditions (\ref{dbc1}, \ref{dbc2}) may be described by a linear system on the transforms. To do this, however, we need to consider the transforms not just at momentum $k$ but also at momentum $\sqrt{1-k^2}$, as in (\ref{trnfd}). This is equivalent to the fact that, in \S4 of \cite{Har9}, we need to consider momenta $(k_1,k_2)$ and $(k_2,k_1)$ for the first and second particles respectively. In this way we see that the linear system induced on (\ref{trnfd}) by (\ref{dbc1}, \ref{dbc2}) is nothing more or less than the linear conditions on the diagonal satisfied by the solutions $G^i$ described in \S4 of \cite{Har9}. \\ 
On each diagonal quadrant we have from (\ref{bsoln})
\begin{eqnarray*}
\psi (x=y^+) & = & \int_{I} \left( \mbox{} - e^{i(k + \sqrt{1-k^2})x}\, \hat{\psi}^{++} + e^{-i(k - \sqrt{1-k^2})x}\, \hat{\psi}^{-+} \right. \\
& & \left. \mbox{} - e^{-i(k + \sqrt{1-k^2})x}\, \hat{\psi}^{--} + e^{i(k - \sqrt{1-k^2})x}\, \hat{\psi}^{+-} \right) dk \\
& = & \int_{\frac{1}{\sqrt{2}}I} \left( \mbox{} - e^{i(k + \sqrt{1-k^2})x} \left[ \hat{\psi}^{++}(k) + \hat{\psi}^{++}\left( \sqrt{1-k^2} \right) \frac{k}{\sqrt{1-k^2}} \right] \right. \\
& & \mbox{} + e^{-i(k - \sqrt{1-k^2})x} \left[ \hat{\psi}^{-+}(k) + \hat{\psi}^{+-}\left( \sqrt{1-k^2} \right) \frac{k}{\sqrt{1-k^2}} \right] \\
& & \mbox{} - e^{-i(k + \sqrt{1-k^2})x} \left[ \hat{\psi}^{--}(k) + \hat{\psi}^{--}\left( \sqrt{1-k^2} \right) \frac{k}{\sqrt{1-k^2}} \right] \\ 
& & \left. \mbox{} + e^{i(k - \sqrt{1-k^2})x} \left[ \hat{\psi}^{+-}(k) + \hat{\psi}^{-+}\left( \sqrt{1-k^2} \right) \frac{k}{\sqrt{1-k^2}} \right] \right) dk 
\end{eqnarray*}
where we split $I$ into $\left[0,\frac{1}{\sqrt{2}}\right]\cup \left[\frac{1}{\sqrt{2}},1\right]$ and change variables in the second interval. Similarly
\begin{eqnarray*}
\psi (x=y^-) & = & \int_{\frac{1}{\sqrt{2}}I} \left( \mbox{} - e^{i(k + \sqrt{1-k^2})x} \left[ \check{\psi}^{++}(k) + \check{\psi}^{++}\left( \sqrt{1-k^2} \right) \frac{k}{\sqrt{1-k^2}} \right] \right. \\
& & \mbox{} + e^{-i(k - \sqrt{1-k^2})x} \left[ \check{\psi}^{-+}(k) + \check{\psi}^{+-}\left( \sqrt{1-k^2} \right) \frac{k}{\sqrt{1-k^2}} \right] \\
& & \mbox{} - e^{-i(k + \sqrt{1-k^2})x} \left[ \check{\psi}^{--}(k) + \check{\psi}^{--}\left( \sqrt{1-k^2} \right) \frac{k}{\sqrt{1-k^2}} \right] \\ 
& & \left. \mbox{} + e^{i(k - \sqrt{1-k^2})x} \left[ \check{\psi}^{+-}(k) + \check{\psi}^{-+}\left( \sqrt{1-k^2} \right) \frac{k}{\sqrt{1-k^2}} \right] \right) dk \, .
\end{eqnarray*}
Since for $k\in\frac{1}{\sqrt{2}}I$, $\mbox{}-k+\sqrt{1-k^2}\in[0,1]$ and $k+\sqrt{1-k^2}\in[1,\sqrt{2}]$, each of the four exponentials in the above integrals takes arguments in disjoint intervals. As a result continuity (\ref{dbc1}) is equivalent to the equality of the coefficients of the exponentials, i.e.
\begin{equation}\label{dbc3}
\check{\psi}^{\sigma \tau}(k) \sqrt{1-k^2} + \check{\psi}^{\tau \sigma}\left( \sqrt{1-k^2} \right) k = \hat{\psi}^{\sigma \tau}(k) \sqrt{1-k^2} + \hat{\psi}^{\tau \sigma}\left( \sqrt{1-k^2} \right) k 
\end{equation}
for $k\in\frac{1}{\sqrt{2}}I$. \\
For the boundary condition (\ref{dbc2}) we again use (\ref{bsoln}) to get
\begin{eqnarray*}
0 & = & \int_{I} \left( e^{i(k + \sqrt{1-k^2})x} \left[ i\left( k - \sqrt{1-k^2} \right) \left[ \mbox{} - \hat{\psi}^{++} + \check{\psi}^{++} \right] + 2c \hat{\psi}^{++} \right] \right. \\
& & \mbox{} + e^{-i(k - \sqrt{1-k^2})x} \left[ i\left( k + \sqrt{1-k^2} \right) \left[ \mbox{} - \hat{\psi}^{-+} + \check{\psi}^{-+} \right] - 2c \hat{\psi}^{-+} \right] \\
& & \mbox{} + e^{-i(k + \sqrt{1-k^2})x} \left[ i\left( k - \sqrt{1-k^2} \right) \left[ \hat{\psi}^{--} - \check{\psi}^{--} \right] + 2c \hat{\psi}^{--} \right] \\ 
& & \left. \mbox{} + e^{i(k - \sqrt{1-k^2})x} \left[ i\left( k + \sqrt{1-k^2} \right) \left[ \hat{\psi}^{+-} - \check{\psi}^{+-} \right] - 2c \hat{\psi}^{+-} \right] \right) dk 
\end{eqnarray*}
on the diagonal quadrants. Splitting $I$ into $\left[0,\frac{1}{\sqrt{2}}\right]\cup \left[\frac{1}{\sqrt{2}},1\right]$ we have the following equalities of transforms
\begin{eqnarray}
0 & = & \left[ \mbox{} - \hat{\psi}^{++} + \check{\psi}^{++} \right] (k) \sqrt{1-k^2} + \left[ \hat{\psi}^{++} - \check{\psi}^{++} \right] \left(\sqrt{1-k^2}\right) k \nonumber \\
& & \mbox{} + 2 c_- \left[ \hat{\psi}^{++} (k) \sqrt{1-k^2} + \hat{\psi}^{++}\left(\sqrt{1-k^2}\right) k \right] \label{dbc4} \\
0 & = & \left[ \hat{\psi}^{--} - \check{\psi}^{--} \right] (k) \sqrt{1-k^2} + \left[ \mbox{} - \hat{\psi}^{--} + \check{\psi}^{--} \right] \left(\sqrt{1-k^2}\right) k \nonumber \\
& & \mbox{} + 2 c_- \left[ \hat{\psi}^{--} (k) \sqrt{1-k^2} + \hat{\psi}^{--}\left(\sqrt{1-k^2}\right) k \right] \label{dbc5} \\
0 & = & \left[ \mbox{} - \hat{\psi}^{-+} + \check{\psi}^{-+} \right] (k) \sqrt{1-k^2} + \left[ \hat{\psi}^{+-} - \check{\psi}^{+-} \right] \left(\sqrt{1-k^2}\right) k \nonumber \\
& & \mbox{} - 2 c_+ \left[ \hat{\psi}^{-+} (k) \sqrt{1-k^2} + \hat{\psi}^{+-}\left(\sqrt{1-k^2}\right) k \right] \label{dbc6} \\
0 & = & \left[ \hat{\psi}^{+-} - \check{\psi}^{+-} \right] (k) \sqrt{1-k^2} + \left[ \mbox{} - \hat{\psi}^{-+} +\check{\psi}^{-+} \right] \left(\sqrt{1-k^2}\right) k \nonumber \\
& & \mbox{} - 2 c_+ \left[ \hat{\psi}^{+-} (k) \sqrt{1-k^2} + \hat{\psi}^{-+}\left(\sqrt{1-k^2}\right) k \right] \label{dbc7} 
\end{eqnarray}
where $c_{\pm}= -ic \left( k \pm \sqrt{1-k^2} \right)^{-1}$ and again $k\in\frac{1}{\sqrt{2}}I$. As we need to consider transforms evaluated at $k$ and $\sqrt{1-k^2}$ these equations are certainly not linear in the variables defined in \S\ref{defel}; however, extending the definition of $\xi$ and $\chi$ to
\begin{equation}\label{trnfd}
\xi_{ij} = \left( \begin{array}{c}
\psi^{++}_{ij}(k) \sqrt{1-k^2} \\ 
\psi^{++}_{ij} \left( \sqrt{1-k^2} \right) k \\ 
\psi^{--}_{ij}(k) \sqrt{1-k^2} \\ 
\psi^{--}_{ij} \left( \sqrt{1-k^2} \right) k \end{array} \right)  \; , \; 
\chi_{ij} = \left( \begin{array}{c}
\psi^{+-}_{ij} (k) \sqrt{1-k^2} \\ 
\psi^{+-}_{ij} \left( \sqrt{1-k^2} \right) k  \\
 \psi^{-+}_{ij} (k) \sqrt{1-k^2} \\ 
\psi^{-+}_{ij} \left( \sqrt{1-k^2} \right) k \end{array} \right) \, , \label{extvar}
\end{equation}
$k\in\frac{1}{\sqrt{2}}I$, we see that (\ref{dbc3}--\ref{dbc7}) can be written as
\begin{equation}\label{eigs}
\hat{\xi}_{ii} = M \check{\xi}_{ii} \; , \;  \hat{\chi}_{ii} = N \check{\chi}_{ii}
\end{equation}
with
\begin{eqnarray*}
M & = & \left( \begin{array}{cccc}
1+c_- & c_- & 0 & 0 \\
-c_- & 1-c_- & 0 & 0 \\
0 & 0 & 1-c_- & -c_-  \\ 
0 & 0 & c_- & 1+c_- \end{array} \right) \\
N & = & \left( \begin{array}{cccc}
1+c_+ & 0 & 0 & c_+ \\ 
0 & 1+c_+ & c_+ & 0  \\ 
0 & -c_+ & 1-c_+ & 0 \\ 
-c_+ & 0 & 0 & 1-c_+ \end{array} \right) \, .
\end{eqnarray*}
Using this new definition of $\xi$, $\chi$ the solutions of the linear system (\ref{defs}) are simply doubled---$\tau$ becomes the permutation $(13)(24)$ on $\C^4$ and the eigenspaces of $\tau$ are now $2$-dimensional. However, since $k\in\frac{1}{\sqrt{2}}I$ in (\ref{extvar}) we still get the same set of solutions: the map (\ref{ontodef}) should be considered as a map from $\L \left( \sqrt{\frac{1}{2}} I \right)\otimes \left( V \oplus V \right)$ where the first copy of $V$ is spanned by elements of the form $F^i \left( k_1, k_2 \right)$ and the second copy of $V$ is spanned by elements of the form $F^i \left( k_2, k_1 \right)$. In this way we see that (\ref{eigs}) is a set of linear equations on $V \oplus V$ with the momenta swopped in the two copies of $V$. But the solution of such a problem is exactly what is described in \S4 of \cite{Har9}. Consequently, $W$ describes the complete solution space of (\ref{defs}, \ref{eigs}) and (\ref{ontosoln}) describes all the eigensolutions of $H_c$ with real momenta.
\subsection{Explicit expressions for eigensolutions of $H_c$}\label{EHc2}
We begin by recalling the definitions of the one and two particle solutions described in \cite{Har9}; the definitions are slightly changed for simplification. \\
The scattering wave solutions $\{\psi^i \}^n_{i=1}$ are defined in exactly the same way as in \cite{Har9}. We also need the solutions
$$
\phi^0 = \frac{1}{2} \sum^n_{j=1} \psi^j \; , \;
\phi^j = \frac{1}{2i} \left( \psi^j - \psi^{j+1} \right) \, ,
$$
where now $j\in\{1,\ldots , n\}$ and $\psi^{n+1}\equiv\psi^1$ (ie. the $\phi^i$ are not linearly independent; however, this allows us to define solutions `mod $n$'). In order to be consistent with the above notation we need new notation for the one particle solution with a discontinuity on the diagonal
$$
\xi (x_i,k) = \left\{ \begin{array}{ll}
\sin (kx_i) & : x_i\in Q_{ii} , \, x_i>y_i \;\; \mbox{or} \;\; x_i\in Q_{ij} , \, i\neq j \\ 
\left( 1 - n \right) \sin (kx_i) & : x_i\in Q_{ii} , \, x_i<y_i
\end{array} \right. 
$$
(with the analogous definition for the second particle ($y$)). The two particle states are then
\begin{eqnarray*}
& \left\{ \Phi^{ij}_{st}  =  \phi^i (x,k_s)\cdot\phi^j (y,k_t) \right\}_{i,j\in\{ 0,\ldots , n\} } & \\
& \left\{ \Psi^{ij}_{st}  =  \psi^i (x,k_s)\cdot\psi^j (y,k_t) \right\}_{i,j\in\{ 1,\ldots , n\} } & \\
& \left\{ \Psi^{i}_{st}  =  \phi^i (x,k_s)\cdot\xi (y,k_t) - \xi (x,k_s)\cdot\phi^i (y,k_t) \right\}_{i\in\{ 1,\ldots , n\} } & \, ,
\end{eqnarray*}
with $(s,t)\in \{ (1,2), (2,1) \}$. \\
In the following we describe the solutions $G^i$ spanning $W$ splitting them into {\em symmetric} and {\em antisymmetric} solutions. Note: here symmetry, antisymmetry, is with respect to the exchange of particles $(x,y)\leftrightarrow(y,x)$.
\begin{enumerate}
\item The $n^2$ antisymmetric solutions
$$
\left\{ \Psi^{ij}_{12}  - \Psi^{ji}_{21} \right\}_{i,j\in\{ 1,\ldots , n\} } \, .
$$
\item The symmetric solutions
\begin{enumerate}
\item with support outside the diagonal
$$
\left\{ \Phi^{ij}_{12}  + \Phi^{ji}_{21} \right\}_{i,j\in\{ 1,\ldots , n\}\, , \, |i-j| {\small{\mod}}n \ge 2 }
$$
of which there are $n^2 - 3n$, and
\item with support on the diagonal
$$
\left\{ \Psi^{i}_{12} - \Psi^{i}_{21} + \frac{n k_1}{c} \left( \Phi^{0i}_{12} + \Phi^{i0}_{21} \right) - \frac{n k_2}{c} \left( \Phi^{0i}_{21} + \Phi^{i0}_{12} \right) \right\}_{i\in\{ 1,\ldots , n\} }
$$
of which there are $n$.
\end{enumerate}
\end{enumerate}
Here $|i-j| \mod n \ge 2$ is to be understood as: the distance between $i$ and $j$, as elements of $\Z/n\Z$, is $\ge 2$. \\
The $2n^2 - 2n$ solutions described above correspond to the solutions described in \S4 of \cite{Har9} in the following way: the $n^2$ antisymmetric solutions correspond to the antisymmetrisation of the solutions from \S4.1 along with the solutions from \S4.2 of \cite{Har9}; the $n^2 - 3n$ symmetric solutions with support outside the diagonal correspond to the symmetrisation of the solutions from \S4.1.1--2 of \cite{Har9}; and the $n$ symmetric solutions with support on the diagonal correspond to the symmetrisation of the solutions from \S4.1.3 and the solutions from \S4.3, of \cite{Har9}.
\section{Conclusion}
As stated above, one of the motivations for studying this problem is that the equivalent problem on the line is completely solvable for an arbitrary number of particles. It is of interest to know how the geometry of a star graph effects this solvability. We see that for two particles the (real momentum) eigensolutions are, somewhat analogous to the line, all essentially described by finite sums of products of one particle solutions (on the configuration space cut along the diagonal). Clearly it would be interesting to consider three or more particles and observe whether the crucial property of the Fourier transforms, viz. the boundary conditions amount to a linear system on some finite dimensional vector space, is satisfied. \\
Another interesting problem is to investigate whether there exist any solutions with non real momenta. We do not attempt a rigorous examination here, rather we merely mention some relevant points. For two particles on the line with an attractive potential ($c<0$) there appears a branch of the continuous spectrum $\left[ \mbox{}-\frac{c^2}{4}, \infty \right)$ of finite multiplicity \cite{Alb:Kur}. For this to appear in the case of the star graph with $n\ge 3$ it should be associated with solutions which have discontinuous derivative on the diagonal, i.e. the solutions described in \S4.3 of \cite{Har9}
$$
G^i = \Psi^{i}_{12} - \Psi^{i}_{21} + \frac{n k_1}{c} \left( \Phi^{0i}_{12} + \Phi^{i0}_{21} \right) - \frac{n k_2}{c} \left( \Phi^{0i}_{21} + \Phi^{i0}_{12} \right) \, .
$$
It can be shown that on the diagonal quadrants this solution has the form
\begin{eqnarray*}
\left. G^i \right|_{Q_{ii}} & = & n \left( \sin k' |x_i-y_i| \sin k (x_i+y_i) - \sin k |x_i-y_i| \sin k' (x_i+y_i) - \mbox{} \right. \\
& & \left. \frac{2k}{c}  \cos k |x_i-y_i| \sin k' (x_i+y_i) + \frac{2k'}{c}  \cos k' |x_i-y_i| \sin k (x_i+y_i) \right)
\end{eqnarray*}
with $k=(k_1+k_2)/2$ and $k'=(k_1-k_2)/2$ momenta along the directions $x_i-y_i$ and $x_i+y_i$. In the case of an attractive potential ($c<0$) the finite multiplicity branches of continuous spectrum appear when one of these momenta, say $k$, is set to $k=\frac{ic}{2}$ so that the solution becomes exponentially decreasing along the direction $x_i-y_i$. Varying the other momentum, as a real parameter, we get a branch of the continuous spectrum starting at $\left( \frac{ic}{2} \right)^2 < 0$.
In the case of $G^i$ this gives
\begin{eqnarray*}
\left. G^i \right|_{Q_{ii}, k=\frac{ic}{2}} & = & i n \left( - e^{\frac{c}{2} |x_i-y_i|} \sin k' (x_i+y_i) + \mbox{} \right. \\
& & \left. \left[ \frac{2k'}{c}  \cos k' |x_i-y_i| + \sin k' |x_i-y_i| \right] \sinh {\textstyle \frac{c}{2}} (x_i+y_i) \right)
\end{eqnarray*}
with the first term exponentially decreasing. However, the second term is exponentially increasing with no apparent (to the author) way of removing it. For this reason it appears probable that such branches of continuous spectrum do not appear for the $n\ge 3$ star graph. A related question is the description of the deficiency elements for the symmetric operator $H_{c=0}$ restricted to functions vanishing along the diagonal $D$, \cite{Alb:Kur}. It would appear that these are related to the basic solutions described \S\ref{defel} (with eigenvalue off the spectrum). Such an operator theoretic description of the problem would be highly desirable and is indeed one of the motivations for this investigation. Such a connection would allow for the application of standard techniques from extension theory to describe eg. the greens function for this problem.
\section*{Acknowledgements}
This paper is dedicated to the memory of Vladimir Geyler. \\
The author is grateful for support from grant LC06002 of the Ministry of Education, Youth and Sport of the Czech Republic.
\section*{Appendix}
Here we define $\{ e_i \}^n_{i=1}$ to be the basis of $\C^n$ in which
$$
P = \frac{1}{n} \left( \begin{array}{ccc}
1 & 1 & \cdots \\
1 & 1 & \cdots \\
\vdots &  & \ddots
\end{array} \right) 
$$
and $\{ f_i \}^n_{i=1}$ to be the basis of $\C^n$ in which
$$
S = 2P - \1 = \left( \begin{array}{cc}
1 & 0 \\
0 & -1 
\end{array} \right) 
$$
where the lower right component of the above matrix is in $M_{n-1}(\C)$. Unless otherwise stated we use the basis $\{ f_i \}$ in this section. \\
Since the transposition $\tau$ commutes with $S$ and $\pi$ we may split (\ref{defs}) into the $\pm 1$ eigenspaces of $\tau$. In this way we see that describing all solutions of (\ref{defs}) amounts to finding the kernels of
$$
P_{\pm} \left( \hat{\chi} , \check{\chi} \right) = \left( \pi^{\perp}\left( \hat{\chi} S \pm S \check{\chi} \right) \, , \, \pi^{\perp}\left( \hat{\chi} - \check{\chi} \right) \right)
$$
on the Hilbert space $M_n(\C)\oplus M_n(\C)$---we consider $M_n(\C)$ instead of $M_n(\C^2)$ since the eigenspaces of $\tau$ are one dimensional. Defining 
\begin{eqnarray*}
Q_{\pm} \left( \hat{\chi} , \check{\chi} \right) & = & \left( \hat{\chi} S \pm S \check{\chi} \, , \, \hat{\chi} - \check{\chi} \right) \, , \\
\Pi^{\perp} \left( \hat{\chi} , \check{\chi} \right) & = & \left( \pi^{\perp}\hat{\chi} \, , \, \pi^{\perp} \check{\chi} \right)
\end{eqnarray*}
along with the subspace
$$
K_{\pm} = \left\{ x\in\ker^{\perp} \left( Q_{\pm} \right) \, ; \, Q_{\pm} x \in \ker \left( \Pi^{\perp} \right) \right\}
$$
it is clear that we have the decomposition
$$
\ker \left( P_{\pm} \right) = \ker \left( Q_{\pm} \right) \oplus K_{\pm} \, .
$$
We see immediately (using $\{ f_i \}$) that
\begin{eqnarray}
\ker \left( Q_{+} \right) & = & \left(  \left( \begin{array}{cc}
0 & \beta \\
b & 0 
\end{array} \right) \, , \, \left( \begin{array}{cc}
0 & \beta \\
b & 0 
\end{array} \right) \right)  \label{kQp} \\
\ker^{\perp} \left( Q_{+} \right) & = & \left(  \left( \begin{array}{cc}
a & \beta \\
b & A 
\end{array} \right) \, , \, \left( \begin{array}{cc}
a' & -\beta \\
-b & A' 
\end{array} \right) \right)  \label{kpQp} \\
\ker \left( Q_{-} \right) & = & \left(  \left( \begin{array}{cc}
a & 0 \\
0 & A 
\end{array} \right) \, , \, \left( \begin{array}{cc}
a & 0 \\
0 & A 
\end{array} \right) \right)  \label{kQm} \\
\ker^{\perp} \left( Q_{-} \right) & = & \left(  \left( \begin{array}{cc}
a & \beta \\
b & A 
\end{array} \right) \, , \, \left( \begin{array}{cc}
-a & \beta' \\
b' & -A 
\end{array} \right) \right) \, . \label{kpQm}
\end{eqnarray}
In particular $\nulty \left( Q_{+} \right)=2(n-1)$ and $\nulty \left( Q_{-} \right)=(n-1)^2+1$. \\
We describe $K_{\pm}$ by first finding $\ker \left( \Pi^{\perp} \right) \cap \ran \left( Q_{\pm} \right)$ and then finding the inverse image with respect to $Q_{\pm}$ in $\ker^{\perp} \left( Q_{\pm} \right)$. For this we need
\begin{eqnarray}
\ran \left( Q_{+} \right) & = & \left(  \left( \begin{array}{cc}
a & \beta \\
b & A 
\end{array} \right) \, , \, \left( \begin{array}{cc}
a' & -\beta \\
b & A' 
\end{array} \right) \right)  \\
\ran \left( Q_{-} \right) & = & \left(  \left( \begin{array}{cc}
a & \beta \\
b & A 
\end{array} \right) \, , \, \left( \begin{array}{cc}
a & \beta' \\
b' & -A 
\end{array} \right) \right) \, .
\end{eqnarray}
We first describe $K_+$. By definition
$$
\ker \left( \Pi^{\perp} \right) = \left( \langle e_i \otimes e_i \rangle^n_{i=1} \, , \,
\langle e_i \otimes e_i \rangle^n_{i=1} \right) \, ,
$$
in particular $\ker \left( \Pi^{\perp} \right)$ consists of symmetric matrices so
$$
\ker \left( \Pi^{\perp} \right) \cap \ran \left( Q_{+} \right) \subset \left(  \left( \begin{array}{cc}
a & 0 \\
0 & A 
\end{array} \right) \, , \, \left( \begin{array}{cc}
a' & 0 \\
0 & A' 
\end{array} \right) \right) \, . 
$$
Consequently, we seek matrices diagonal in the $\{ e_i \}$ basis which commute with $P$; but the only such matrices are the scalar matrices so 
$$
\ker \left( \Pi^{\perp} \right) \cap \ran \left( Q_{+} \right) = \left( 2a\1 , 2a'\1 \right) \, .
$$
The preimage of this in $\ker^{\perp} \left( Q_{+} \right)$, from (\ref{kpQp}), is
$$
\left(  \left( \begin{array}{cc}
a+a' & 0 \\
0 & -(a-a')\1 
\end{array} \right) \, , \, \left( \begin{array}{cc}
a-a' & 0 \\
0 & -(a+a')\1 
\end{array} \right) \right) \, ,
$$
so that $\dim \left( K_{+} \right) = 2$. \\
For $K_{-}$ we have
$$
\ker \left( \Pi^{\perp} \right) \cap \ran \left( Q_{-} \right) \subset \left(  \left( \begin{array}{cc}
a & b \\
b & A 
\end{array} \right) \, , \, \left( \begin{array}{cc}
a & b' \\
b' & -A 
\end{array} \right) \right)
$$
again using the fact that $\ker \left( \Pi^{\perp} \right)$ consists of symmetric matrices. Here the repeated $b$ ($b'$) are understood to be transposes of one another. We should seek solutions of the form
$$
\chi = \sum_i c_i\, e_i \otimes e_i = \left( \begin{array}{cc}
a & b \\
b & A 
\end{array} \right) \, , \, \chi' = \sum_i c'_i\, e_i \otimes e_i = \left( \begin{array}{cc}
a & b' \\
b' & -A 
\end{array} \right) \, ,
$$
so that 
$$
\chi'' = \chi + \chi' = \sum_i c''_i\, e_i \otimes e_i = \left( \begin{array}{cc}
2a & b+b' \\
b+b' & 0 
\end{array} \right) 
$$
with $c''_i = c_i + c'_i$. By construction
\begin{eqnarray*}
0 & = & P^{\perp} \chi'' P^{\perp} \\
& = & \left[ \1 - \frac{1}{n} \left( \sum_i e_i \right) \otimes \left( \sum_j e_j \right) \right] \left[ \sum_i c''_i\, e_i \otimes e_i \right] \times \mbox{} \\
& & \left[ \1 - \frac{1}{n} \left( \sum_i e_i \right) \otimes \left( \sum_j e_j \right) \right] \\
& = & \sum_i c''_i\, e_i \otimes e_i - \frac{1}{n} \sum_{i,j} e_i \otimes e_j \left[ c''_i + c''_j - \frac{1}{n} \sum_k c''_k \right] \, .
\end{eqnarray*}
This implies that $c''_i = c_i + c'_i = 0$ so that
\begin{equation}\label{imageKm}
\ker \left( \Pi^{\perp} \right) \cap \ran \left( Q_{-} \right) \subset \left(  \left( \begin{array}{cc}
0 & b \\
b & A 
\end{array} \right) \, , \, \left( \begin{array}{cc}
0 & -b \\
-b & -A 
\end{array} \right) \right)
\end{equation}
which in turn gives
\begin{eqnarray*}
0 & = & P \chi P \\
& = & \frac{1}{n^2} \sum_k c_k \sum_{i,j} e_i \otimes e_j 
\end{eqnarray*}
or $\sum_k c_k=0$. In this way we see that $\ker \left( \Pi^{\perp} \right) \cap \ran \left( Q_{-} \right)$ is $n-1$ dimensional and is spanned by
\begin{equation}\label{Kmsoln}
\left( \sum_i c_i\, e_i \otimes e_i , - \sum_i c_i\, e_i \otimes e_i \right) \; , \; \sum_k c_k=0 \, . 
\end{equation}
We describe the inverse image: using (\ref{kpQm}) we see that the inverse image of (\ref{imageKm}) in $\ker^{\perp} \left( Q_{-} \right)$ is
$$
K_- \subset \left(  \left( \begin{array}{cc}
0 & b \\
0 & A/2 
\end{array} \right) \, , \, \left( \begin{array}{cc}
0 & 0 \\
-b & -A/2 
\end{array} \right) \right) \, .
$$
Suppose we have solutions of the form (\ref{Kmsoln}) then we claim that
$$
\frac{1}{n} \sum_i c_i\, e_i \otimes \sum_j e_j =  \left( \begin{array}{cc}
0 & 0 \\
b & 0 
\end{array} \right)
$$
This is seen to be true since
$$
\left( \begin{array}{c}
0 \\
b  
\end{array} \right) = \left( \begin{array}{cc}
0 & b \\
b & A 
\end{array} \right) f_1 = \left( \sum_i c_i\, e_i \otimes e_i \right) f_1 = \frac{1}{\sqrt{n}} \sum_i c_i\, e_i \, .
$$
Consequently, the inverse image of $Q_{-}$ in $\ker^{\perp} \left( Q_{-} \right)$ is spanned by
\begin{eqnarray*}
\left( \sum_i c_i\, e_i \otimes e_i + \frac{1}{n} \sum_i e_i \otimes \sum_i c_j\, e_j + \frac{1}{n} \sum_i c_i\, e_i \otimes \sum_j e_j \, , \right. \\
\left. \mbox{} - \sum_i c_i\, e_i \otimes e_i + \frac{1}{n} \sum_i e_i \otimes \sum_i c_j\, e_j + \frac{1}{n} \sum_i c_i\, e_i \otimes \sum_j e_j \right) 
\end{eqnarray*}
with $\sum_k c_k=0$. In particular it also has dimension $n-1$. \\
In summary, the solutions of (\ref{defs}) split into four subspaces: $\ker \left( Q_{-} \right)$, $\ker \left( Q_{+} \right)$, $K_{-}$ and $K_{+}$ of dimensions $(n-1)^2+1$, $2(n-1)$, $n-1$ and $2$ respectively.

\end{document}